\documentclass[twocolumn,english,prl,amsmath,amssymb,preprintnumbers]{revtex4}
\usepackage[T1]{fontenc}
\usepackage[latin9]{inputenc}
\usepackage{amstext}
\usepackage{amsmath}
\usepackage{amsfonts}
\usepackage{amssymb}
\usepackage{mathtools}
\usepackage{graphicx}
\usepackage{rotating}
\usepackage {bm}
\usepackage[section]{placeins}
\usepackage{color}
\usepackage{multirow}
\usepackage{cellspace}
\makeatletter
\@ifundefined{textcolor}{}
{%
 \definecolor{BLACK}{gray}{0}
 \definecolor{WHITE}{gray}{1}
 \definecolor{RED}{rgb}{1,0,0}
 \definecolor{GREEN}{rgb}{0,1,0}
 \definecolor{BLUE}{rgb}{0,0,1}
 \definecolor{CYAN}{cmyk}{1,0,0,0}
 \definecolor{MAGENTA}{cmyk}{0,1,0,0}
 \definecolor{YELLOW}{cmyk}{0,0,1,0}
}

\def\kF{k_{\text{F}}}
\def\vF{v_{\text{F}}}
\def\NF{N_{\text{F}}}

\def\chis{\chi_{\text{s}}}

\def\sgn{{\text{sgn\,}}}

\def\be{\begin{equation}}
\def\ee{\end{equation}}
\def\bea{\begin{eqnarray}}
\def\eea{\end{eqnarray}}
\def\bse{\begin{subequations}}
\def\ese{\end{subequations}}

\def\chis{\chi_{\text s}}

\def\Gammat{\Gamma_{\text t}}

\usepackage{dcolumn}
\usepackage{bm}
\usepackage{amsfonts}
\draft

\makeatother

\usepackage{babel}
\begin{document}
%

\title{Quantum Ferromagnetic Transition in Clean Dirac Metals}

\author{T. R. Kirkpatrick$^{1}$ and  D. Belitz$^{2,3}$}

\affiliation{$^{1}$ Institute for Physical Science and Technology, University of Maryland, College Park, MD 20742, USA\\
                 $^{2}$ Department of Physics and Institute of Theoretical Science, University of Oregon, Eugene, OR 97403, USA\\
                 $^{3}$ Materials Science Institute, University of Oregon, Eugene, OR 97403, USA
                  }

\date{\today}
\begin{abstract}
The ferromagnetic quantum phase transition in clean metals with a negligible spin-orbit interaction is known to be first order due to a coupling of the
magnetization to soft fermionic particle-hole excitations. A spin-orbit interaction gives these excitations a mass, suggesting the existence of a
ferromagnetic quantum critical point in metals with a strong spin-orbit interaction. We show that this expectation is not borne out in a large class
of materials with a Dirac spectrum, since the chirality degree of freedom leads to new soft modes that again render the transition first order.
\end{abstract}
%
%
\maketitle

Solids in which a strong spin-orbit coupling leads to the linear crossing of doubly degenerate bands \cite{Herring_1937, Abrikosov_Beneslavskii_1970}
have attracted much attention in recent years, most of which has focused on the topological properties of such materials \cite{Armitage_Mele_Vishwanath_2018}. 
If the chemical potential $\mu$ is tuned to the crossing point one has a semimetal, if the crossing point is gapped out and $\mu$ 
lies in the gap one has a topological insulator. If $\mu$ lies within the conduction band one has a true metal, but the crossing point, 
whether gapped out or not, still leads to unusual properties. In all of these cases the effective Hamiltonian in the vicinity of the crossing point is reminiscent
of a massless (for a gapless system) or massive (for a gapped one) Dirac Hamiltonian. We will be interested in the case of a true
metal, which we will refer to as a Dirac metal (DM). Specifically, we will investigate the nature of the
quantum phase transition (QPT) from a paramagnetic DM to a ferromagnetic one. This question is of particular
interest since some of the recently found Dirac materials are magnetic \cite{Liu_et_al_2017}.

In clean metals with a negligibly weak spin-orbit interaction the ferromagnetic QPT is discontinuous,
or first order, in all spatial dimensions $d>1$ \cite{Belitz_Kirkpatrick_Vojta_1997, Belitz_Kirkpatrick_Vojta_1999, Kirkpatrick_Belitz_2012b, Brando_et_al_2016a}.
At nonzero temperature ($T$) there is a tricritical point  on the phase boundary separating a line of second-order transitions at high $T$
from a line of first-order transitions at low $T$. In an external magnetic field there are tricritical wings that end in critical points at the
wing tips \cite{Belitz_Kirkpatrick_Rollbuehler_2005}. Numerous experiments have confirmed these predictions \cite{Brando_et_al_2016a}. 

A spin-orbit interaction splits the Fermi surface, which is expected to suppress the
soft modes that cause the first-order transition in ordinary metals. A fundamental question then is whether a strong spin-orbit interaction restores a
ferromagnetic quantum critical point in a DM, or whether there is another universal mechanism driving the transition first order that is
operative even in a DM. We will show that, generically, the answer is the latter.

The first-order mechanism hinges on the nature of electronic soft modes that couple to the magnetization. 
It is useful to first discuss these soft modes in an ordinary metal. To make the salient point it suffices to
consider the Green function for a Fermi gas, which can be written



\bse
\label{eqs:1}
\be
G_k = \frac{1}{2}\,\sum_{\sigma=\pm} G_k^{\sigma}\,M(\sigma\hat{\bm h})\ .
\label{eq:1a}
\ee
Here $\sigma=\pm$ is the spin projection, and
\bea
G_k^{\sigma} &=& 1/(i\omega_n - \xi_{\bm k} + \sigma h)\ ,
\label{eq:1b}\\
M(\hat{\bm h}) &=& \sigma_0 - {\bm\sigma}\cdot\hat{\bm h}\ .
\label{eq:1c}
\eea
\ese
Here $k = (\omega_n, {\bm k})$ comprises a fermionic Matsubara frequency $\omega_n$ and a wave vector ${\bm k}$, and $\xi_{\bm k} = \epsilon_{\bm k} - \mu$ 
with $\epsilon_{\bm k}$ the single-particle energy. $\sigma_0$ is the $2\times 2$ unit matrix, and $\bm\sigma = (\sigma_1,\sigma_2,\sigma_3)$ 
are the Pauli matrices. ${\bm h}$ is a magnetic field with amplitude $h=\vert{\bm h}\vert$, and $\hat{\bm h} = {\bm h}/h$. Let $q = (\Omega_n,{\bm q})$ with $\Omega_n$ a bosonic  Matsubara frequency, and consider wave vector convolutions of the form
\bea
\varphi_{\sigma,\sigma'}({\bm q},i\Omega_n;i\omega_m) &=& \frac{1}{V} \sum_{\bm k} G_k^{\sigma}\,G_{k+q}^{\sigma'}
\nonumber\\
&&\hskip -100pt = \int \frac{d\Omega_{\bm k}}{4\pi}\,\frac{-2\pi i \NF\,\sgn(\omega_m)\,\Theta(-\omega_m(\omega_m + \Omega_n))}{i\Omega_n - \vF {\hat{\bm k}}\cdot{\bm q} 
       + (\sigma' - \sigma)h} \ .
\label{eq:2}
\eea
Here $\Omega_{\bm k}$ is the solid angle with respect to the wave vector ${\bm k}$, and the radial 
integration has been done in the well-known approximation that captures the part of the integral that is singular in the limit 
${\bm q}, \Omega_n \to 0$ \cite{Abrikosov_Gorkov_Dzyaloshinski_1963}.

Important features of this result are: (1) The correlations described by these convolutions are soft modes that scale as $1/q$. 
They are ballistic in nature, i.e., the frequency scales as the wave number. (2) The modes are soft only if the
 frequencies of the two Green functions have opposite signs. (3) The modes with $\sigma = \sigma'$ (the spin singlet and the longitudinal part of the spin triplet) are soft
independent of ${\bm h}$, whereas those with $\sigma\neq\sigma'$ (the transverse parts of the spin triplet) acquire a mass if $h>0$. (4) Convolutions
of $n>2$ Green functions scale as $1/q^{n-1}$ provided the $n$ frequencies do not all have the same sign. (5) Spin
conservation is important for this structure. In particular, a spin-orbit interaction is expected to render massive the modes with $\sigma\neq\sigma'$.

In a Fermi liquid, i.e., in the presence of an electron-electron interaction, these conclusions remain valid provided the product of $n$ Green functions
is replaced by an appropriate $2n$-point correlation function that factorizes into the product of Green functions in the noninteracting limit. The simplest
argument for this relies on Landau Fermi-liquid theory which states that the quasi-particle states of the interacting system are adiabatically related to those of
the noninteracting one \cite{Abrikosov_Gorkov_Dzyaloshinski_1963}. A different argument is that the soft modes are the Goldstone modes related to
a symmetry between retarded and advanced degrees of freedom that is spontaneously broken with or
without interactions as long as there is a nonvanishing density of states \cite{Wegner_1979, Belitz_Kirkpatrick_2012a, Kirkpatrick_Belitz_2019}. 


These soft modes are responsible for various nonanalyticities of observables in Fermi liquids that have been known for a long time.
In particular, the spin susceptibility $\chi_s$ is a nonanalytic function of the wave number, temperature,
or magnetic field \cite{Carneiro_Pethick_1977, Belitz_Kirkpatrick_Vojta_1997, Betouras_Efremov_Chubukov_2005}. 
For our purposes we focus on the magnetic field dependence, which in a 3-d system at $T=0$ has the form
\be
\chis(h\to 0) = {\chis}^{(0)} + {\chis}^{(2)}\,h^2 \ln(1/h) + O(h^2)\ .
\label{eq:3}
\ee
The coefficient $\chis^{(2)}$ is positive \cite{positivity_footnote} and for weak electron-electron interactions it is proportional to ${\Gammat}^2$, with $\Gammat$ 
a spin-triplet interaction amplitude \cite{interaction_footnote}. The logarithmic $h$-dependence reflects the fact that the only soft modes that couple to $\chis$
are the two transverse spin-triplet channels ($\sigma' \neq \sigma$ in Eq.~(\ref{eq:2})) that are cut off by $h\neq 0$.

As mentioned above, a spin-orbit coupling is expected to cut off this singularity and make $\chis$ an analytic function of $h$. We will confirm this expectation
below. 

The nonanalyticity shown in Eq.~(\ref{eq:3}) has profound consequences for the ferromagnetic QPT in a clean metal. Since the soft modes couple to
the spin density  ${\bm n}_{\text s}$, they also couple to the magnetization ${\bm m}$ as the latter couples to the former via a Zeeman contribution to
the action,
\be
S_{\text Z} = c \int d{\bm x}\,d\tau\ {\bm m}({\bm x},\tau)\cdot{\bm n}_{\text s}({\bm x},\tau)
\label{eq:4}
\ee
where $c$ is a coupling constant and $\bm x$ and $\tau$ denote real-space position and imaginary time, respectively. 
Any order-parameter theory in terms of the magnetization therefore will contain correlation functions of  ${\bm n}_{\text s}$ 
in its vertices, and in particular incorporate the nonanalytic behavior of $\chis$. Within a renormalized mean-field theory 
this leads to a free energy density \cite{Belitz_Kirkpatrick_Vojta_1997, Belitz_Kirkpatrick_Vojta_1999, Kirkpatrick_Belitz_2012b,
Brando_et_al_2016a}
\be
f = t\,m^2 + {\tilde u}\,m^2 \ln(1/m) + u\,m^4\ .
\label{eq:5}
\ee
$t$, $\tilde u$, and $u$ are Landau parameters. Importantly, $\chis^{(2)}>0$ implies $\tilde u > 0$. As a result, the QPT described by Eq.~(\ref{eq:5})
is necessarily first order. The fermionic fluctuations that lead to the nonanalytic term in Eq.~(\ref{eq:5}) thus invalidate Hertz's conclusion
\cite{Hertz_1976} that the ferromagnetic QPT is second order with mean-field critical behavior. The first-order nature of the transition described by
Eq.~(\ref{eq:5}) is in excellent agreement with experiment \cite{Brando_et_al_2016a}. 

The central question of the current paper is: What will a spin-orbit interaction do to these effects? As mentioned above, it is
expected to make massive the underlying soft modes, which suggests that a sufficiently strong spin-orbit interaction will restore a ferromagnetic
quantum critical point. We will show that this conclusion is not correct for DMs as defined above. While the spin-triplet modes that are
soft in a Landau Fermi liquid do indeed become massive,  the chiral degree of freedom in a DM leads to a new class of soft modes that
restore the nonanalyticity of $\chis$. We thus predict that, surprisingly, the ferromagnetic QPT in a DM is first order,
just as it is in ordinary metals. A possible exception are DMs where a symmetry enforces a gapless spectrum, see below. These predictions are
relevant for a large class of Dirac materials, some of which are magnetic \cite{Liu_et_al_2017}.

We consider systems in which the spin-orbit interaction causes a linear crossing of two bands via a term $\pm{\bm k}\cdot{\bm\sigma}$
in the single-particle Hamiltonian \cite{Abrikosov_Beneslavskii_1970}. Within a simple isotropic model, the most general Hamiltonian that
respects both time reversal (in the absence of a magnetic field) and spatial inversion symmetry reads \cite{Zhang_et_al_2009}
\bea
H_0 &=& (\epsilon_{\bm k} - \mu)(\pi_0\otimes\sigma_0) + v(\pi_3\otimes{\bm\sigma})\cdot{\bm k} +\Delta(\pi_1\otimes\sigma_0) 
\nonumber\\
&& \hskip 0pt - h(\pi_0\otimes\sigma_3) \ .
\label{eq:6}
\eea
The Pauli matrices $\bm\sigma$  represent the physical spin degree of freedom. A related class of models where $\bm\sigma$ represents
 a pseudo-spin degree of freedom, as is the case in graphene \cite{Kotov_et_al_2012}, behave very differently with respect to the properties
we will discuss. The $\pi_i$ in Eq.~(\ref{eq:6}) are a second set of Pauli matrices that represent the chirality 
degree of freedom necessary to insure inversion symmetry.
$v$ is a characteristic velocity that measures the strength of the spin-orbit interaction, and $\Delta$ 
produces a gap in the single-particle spectrum. The magnetic field ${\bm h}$ has been chosen to point in the 3-direction. 

\begin{figure}[t]
\includegraphics[width=8.5cm]{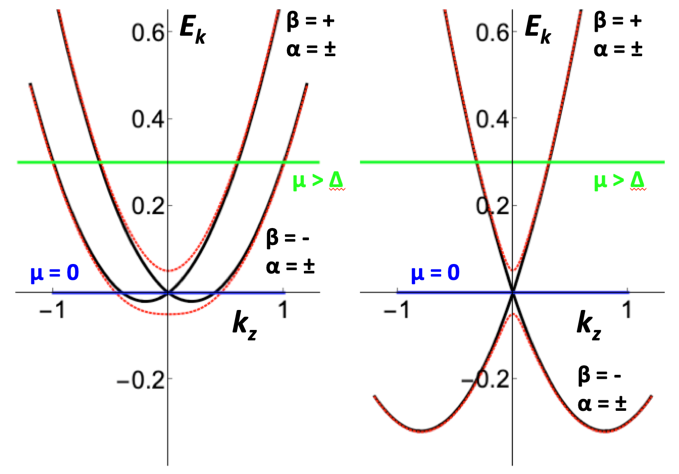}
\caption{Single-particle spectra for $h=0$, $v=0.2$ (left) and $0.8$ (right) in atomic units. Solid black lines are for
              $\Delta=0$, dotted red lines for $\Delta = 0.05$. The up-cone ($\beta=+1$) and down-cone ($\beta=-1$) branches are two-fold
              degenerate each with respect to the chirality index $\alpha=\pm1$. A magnetic field $h>0$ splits this degeneracy.
              The horizontal lines indicate the chemical potential. For $\mu \gg \Delta$ 
              (green horizontal lines) the system is a Dirac metal, for $\mu = 0$ (blue horizontal lines) it is a semimetal (for $\Delta = 0$)
              or an insulator (for $\Delta > 0$).}
\label{fig:1}
\end{figure}

The single-particle spectrum $E_{\bm k}$ is obtained by finding the eigenvalues of $H_0$. We introduce an
atomic-scale momentum $p_0$, velocity $v_0 = p_0/2m$, and energy $E_0 = p_0^2/2m$, with $m$ the effective electron mass, and measure
$E_{\bm k}$, $\Delta$, and $h$ in units of $E_0$, $v$ in units of $v_0$, and ${\bm k}$ in units of $p_0$. Figure~\ref{fig:1} shows
the spectrum for $k_x=k_y=0$, $h=0$, and two values of $v$. We are interested in true metals, where $\mu > \Delta \geq 0$. 
For a given value of $v$ the spectrum near the Fermi surface is then
qualitatively independent of whether $\Delta=0$ or $\Delta>0$. However, the nature of the ferromagnetic QPT may depend on whether or not the
spectrum is gapped, see below. 

%
%
The Fermi liquid we are interested in is governed by $H_0$ plus all interaction amplitudes that are compatible with
the symmetry of $H_0$. Since we are dealing with a true metal, screening works and the interactions can be localized in both space
and time. Only amplitudes in the spin-triplet channel contribute to the nonanalytic behavior of the spin susceptibility. Furthermore, for reasons that
will become clear later, only scattering processes that mix chiralities are relevant for our purposes. There are two interaction processes that
are consistent with these criteria, viz. 
\be
S_{\text{int}} = \frac{T}{2V} \sum_q\!  \sum_{\alpha\neq\alpha'}  \left( \Gamma_{{\text t},3} {\bm\Psi}^{\alpha\alpha'}_q \cdot{\bm\Psi}^{\alpha'\alpha}_{-q}
                         +  \Gamma_{{\text t},4} {\bm\Psi}^{\alpha\alpha'}_q \cdot {\bm\Psi}^{\alpha\alpha'}_{-q} \right)
\label{eq:7}                         
\ee
Here ${\bm\Psi}^{\alpha\alpha'}_q = \sum_k {\bar\psi}^{\alpha}(k){\bm\sigma}\psi^{\alpha'}(k-q)$ with $\bar\psi$ and $\psi$ fermionic spinor fields.
$\Gamma_{\text{t,4}}$ was not considered in Ref.~\onlinecite{Kirkpatrick_Belitz_2019}. It breaks the conservation of the number of particles with a given chirality.
The gap $\Delta$ in $H_0$ breaks the same symmetry, while the other terms in $H_0$ respect it. $\Gamma_{\text{t,4}}$ is therefore
not allowed in systems where $\Delta=0$ due to a crystal symmetry, as is the case in some materials  \cite{Wang_et_al_2013, Neupane_et_al_2014}.

We now discuss the soft modes in the chiral Fermi liquid described by Eqs.~(\ref{eq:6}, \ref{eq:7}) that are analogous to Eq.~(\ref{eq:2}).
The Green function for the Hamiltonian $H_0$ can be written as a generalization of Eq.~(\ref{eq:1a}):
\be
G_k = \frac{1}{2} \sum_{\alpha,\beta=\pm} F_k^{\alpha\beta}\,M_{\alpha\beta}(\hat{\bm k})
\label{eq:8}
\ee
$F$ and $M$ for arbitrary parameter values are complicated, but simplify in the limits $\Delta=0$ and $\Delta\gg v\kF$,
respectively, where $\kF$ is the Fermi wave number. We find
\begin{widetext}
\bea
\Delta = 0&:& \quad F_k^{\alpha\beta} = 1/(i\omega_n - \xi_{\bm k} + \beta\vert v{\bm k} - \alpha {\bm h}\vert)\quad,\quad
   M_{\alpha\beta}(\hat{\bm k}) = 
   \left(\pi_0 + \alpha \pi_3\right) \otimes M(\alpha\beta\hat{\bm k})/2\ ,
\label{eq:9}\\
\Delta \gg v\kF&:& \quad F_k^{\alpha\beta} = 1/(i\omega_n - \xi_{\bm k} + \beta\vert\Delta - \alpha {\bm h}\vert)\ \quad,\quad
   M_{\alpha\beta}(\hat{\bm k}) = 
   \left(\pi_0 - \beta \pi_1\right)\otimes \left(\sigma_0 - \alpha\beta\sigma_3\right)/2\ .
\label{eq:10}
\eea
with $M(\hat{\bm k})$ from Eq.~(\ref{eq:1c}). The generalization of the convolution $\varphi$ in Eq.~(\ref{eq:2}) thus reads
\bse
\label{eqs:11}
\be
\varphi_{\beta_1,\beta_2}^{\alpha_1\alpha_2}({\bm q},i\Omega_n;i\omega_m) = \frac{1}{V}\sum_{\bm k} F_k^{\alpha_1 \beta_1} F_{k+q}^{\alpha_2 \beta_2}
= \int \frac{d\Omega_{\bm k}}{4\pi} \frac{-2\pi i \NF\, \sgn(\omega_m) \, \Theta(-\omega_m(\omega_m + \Omega_n))}{N({\bm q},i\Omega_n;i\omega_m) } 
\label{eq:11a}
\ee
where
\be
N({\bm q},i\Omega_n;i\omega_m) = \begin{cases} i\Omega_n - \vF{\hat{\bm k}}\cdot{\bm q} + \beta_2\vert v(\kF \hat{\bm k} + {\bm q}) - \alpha_2 {\bm h}\vert 
     - \beta_1\vert v\kF \hat{\bm k} - \alpha_1{\bm h}\vert \quad \text{for}\quad \Delta=0\\
i\Omega_n - \vF{\hat{\bm k}}\cdot{\bm q} + \beta_2\vert\Delta - \alpha_2 h\vert - \beta_1\vert\Delta - \alpha_1 h\vert \hskip 66pt \text{for}\quad \Delta \gg v\kF
\end{cases}
\label{eq:11b}
\ee
\ese
\end{widetext}

For $v = \Delta = 0$ the cone index $\beta$ reverts to the spin projection index $\sigma$ and we recover Eq.~(\ref{eq:2}).
For a chiral Fermi gas the modes with $\beta_1 \neq \beta_2$ acquire a mass due to $v$, or $\Delta$, or both. That is, the modes that  led to the
nonanalyticity of $\chis$ in an ordinary Fermi gas, Eq.~(\ref{eq:3}), cannot do so in a chiral one. 
However, for $\beta_1 = \beta_2$ the modes
in Eq.~(\ref{eq:11a}) are soft, and for $\alpha_1 \neq \alpha_2$ these soft modes are cut off by $h$ \cite{cutoff_footnote}. 
These conclusions remain valid in a chiral Fermi liquid for the same reasons an in an ordinary Fermi liquid. 
The chirality thus restores the possibility of a nonanalytic $\chis$ that has the form 
shown in Eq.~(\ref{eq:3}), and provide a new mechanism for the ferromagnetic QPT to be first order. The details depend on the 
parameter values in $H_0$, as we will now discuss.

We first consider the generic case of a gapped DM. The qualitative features are captured by the limit $\Delta \gg v\kF$, which greatly
simplifies the calculations. For $\beta_1 = \beta_2$ the modes are soft, and for $\alpha_1 \neq \alpha_2$ the singularity is cut off by
a magnetic field. The problem then maps onto the ordinary Fermi liquid, with the chirality degree of freedom in the chiral case
playing the role of the spin projection in the non-chiral one. $\chis$ is given by Eq.~(\ref{eq:3}) with both $\Gamma_{{\text t},3}$
and $\Gamma_{{\text t},4}$ contributing to the coefficient $\chis^{(2)}$. The arguments that lead to Eq.~(\ref{eq:5}) then carry
through and the ferromagnetic QPT is first order.

The case of a gapless DM is more complicated. Let us assume that $\Delta=0$ due to a lattice symmetry, which
implies $\Gamma_{{\text t},4} = 0$ as well. This case was considered in Ref.~\cite{Kirkpatrick_Belitz_2019}. To second
order in $\Gamma_{{\text t},3}$, $\chis$ is given by integrals over products of two convolutions that are $3$-$F$ generalizations of
Eqs.~(\ref{eqs:11}). Within each convolution the cone indices $\beta$ must all be the same, as explained above, but the 
overall integral is nonzero only if the two convolutions correspond to different cones. 
%
This is because for intra-cone scattering the
magnetic field ${\bm h}$ can be eliminated by a shift of the hydrodynamic wave vector ${\bm q}$, and hence these processes
cannot contribute to a nonanalytic $h$-dependence of $\chis$. This has important consequences: If $v$ is sufficiently large,
then only the upper cone contributes to the Fermi surface, see Fig.~\ref{fig:1}. This suggests that
if $v$ is large enough, $\chis$ is an analytic function of $h$ and the ferromagnetic QPT is second order
since the soft modes cannot couple to the order parameter. In this case Hertz
theory \cite{Hertz_1976} is expected to apply, and the critical behavior will be mean-field-like with the dynamics affected by
a dangerous irrelevant variable \cite{Millis_1993}. For
smaller values of $v$ both cones contribute to the Fermi surface, see Fig.~\ref{fig:1}, $\chis$ is
nonanalytic, and the transition is first order.

Finally, if $\Delta = 0$ due to fine tuning (e.g., by doping a gapped DM), then there is no reason for  $\Gamma_{{\text t},4}$
to be zero, and the situation is different again: The integral that contributes to the nonanalyticity of $\chis$ now is nonzero
even if the cone indices of the two convolutions are the same. In this respect  $\Gamma_{{\text t},4} \neq 0$ has the same
effect as $\Delta \neq 0$. This was to be expected since the two terms break the same gauge symmetry, see
the remark after Eq.~(\ref{eq:7}). As a result, one cone contributing to the Fermi surface suffices for producing a 
nonanalyticity in $\chis$, and the ferromagnetic QPT is first order even for $v$ so large that only the up-cone
contributes to the Fermi surface. 


We conclude with several remarks. 
(1) While we have used many-body diagrammatic techniques for the calculations, the selection of the diagrams
is informed by an underlying effective field theory as explained in Ref.~\cite{Kirkpatrick_Belitz_2019}. We have
performed the calculation to one-loop order and in addition have restricted ourselves to second order in the
electron-electron interaction. Renormalization-group arguments within the field theory show that $\chis$
{\em must} scale with $h$ as shown in Eq.~(\ref{eq:3}) in $d=3$, and as $\chis \propto \text{const.} + h^{d-1}$
in generic dimensions $1<d<3$. This result is therefore exact as far as the functional form of the nonanalyticity 
is concerned, and the perturbative calculation merely confirms that the prefactor is nonzero. 

(2) The preceding considerations imply that our results will not qualitatively change if one goes to higher order in the interaction expansion. 
For generic (i.e. gapped) DMs they cannot change by going to higher order in the loop expansion either. However, one open question is 
whether effects at higher loop order will restore the missing coupling in a DM that is gapless by symmetry and has a large spin-orbit coupling 
$v$, and thus render the QPT in such systems first order as well. 


(3) 
For the velocity $v$, which measures the strength of
the spin-orbit interaction, there are three relevant regimes: (i) $v\kF$ is small compared to the 
discontinuity of the magnetization at the first-order QPT for $v=0$, both measured in atomic units.
While $v$ changes the soft-mode structure, the effect is too small to override the first-order mechanism
that is operative for $v=0$. $v$ then is negligible and we have the case of an ordinary metal as
discussed before \cite{Belitz_Kirkpatrick_Vojta_1999, Brando_et_al_2016a}. 
(ii) $v$ is not negligible, but still small enough for both cone branches to contribute to the Fermi surface,
see Fig.~\ref{fig:1}. 
The soft-mode structure is now qualitatively different from an ordinary metal, the system is a DM, and the chiral soft modes are
crucial for the conclusion that the QPT is still first order. This holds irrespective of whether the DM is 
gapped or not. 
(iii) $v$ is so large that only the up-cone branch contributes to the Fermi surface, 
Fig.~\ref{fig:1}. 
It now becomes important to distinguish between: (a) $\Delta \gg v\kF$. Then
$v\kF = v\sqrt{2m(\mu-\Delta)}$, so this case can be realized if $\mu\agt\Delta$, even if $v\alt v_0$. In this limit the
analysis of the QPT maps onto the ordinary-metal case, even though the soft-mode structure is physically
very different, and the QPT is again first order. (b) $\Delta \ll v\kF$. 
Then $v \kF = \sqrt{2mv^2\mu}$ if $mv^2\ll\mu$, or $v\kF=\mu$ if
$mv^2\gg\mu$. In either case $\Delta < \mu$, as required for a metal. In this limit the QPT
is second order in our one-loop calculation if $\Gamma_{{\text t},4}=0$, but first order if
$\Gamma_{{\text t},4} \neq 0$.

(4) 
The graphene-type models that are given by Eq.~(\ref{eq:6}) with $\bm\sigma$ representing a
pseudo-spin rather than the physical spin behave very differently. In such systems, the soft-mode spectrum
is the same as in ordinary metals and the ferromagnetic QPT is always first order. 

(5) An alternative consequence of the
nonanalytic $\chis$ is a QPT to an inhomogeneous magnetic state, which may compete with
the first-order transition to a homogeneous ferromagnet. For ordinary metals this possibility was suggested
in Ref.~\cite{Belitz_Kirkpatrick_Vojta_1997} and studied in detail in Refs.~\cite{Conduit_Green_Simons_2009,
Karahasanovic_Kruger_Green_2012}. An analogous investigation is needed for DMs. 


This work was initiated at the Telluride Science Research Center (TSRC). We thank George de Coster for discussions.


\begin{thebibliography}{26}
\expandafter\ifx\csname natexlab\endcsname\relax\def\natexlab#1{#1}\fi
\expandafter\ifx\csname bibnamefont\endcsname\relax
  \def\bibnamefont#1{#1}\fi
\expandafter\ifx\csname bibfnamefont\endcsname\relax
  \def\bibfnamefont#1{#1}\fi
\expandafter\ifx\csname citenamefont\endcsname\relax
  \def\citenamefont#1{#1}\fi
\expandafter\ifx\csname url\endcsname\relax
  \def\url#1{\texttt{#1}}\fi
\expandafter\ifx\csname urlprefix\endcsname\relax\def\urlprefix{URL }\fi
\providecommand{\bibinfo}[2]{#2}
\providecommand{\eprint}[2][]{\url{#2}}

\bibitem[{\citenamefont{Herring}(1937)}]{Herring_1937}
\bibinfo{author}{\bibfnamefont{C.}~\bibnamefont{Herring}},
  \bibinfo{journal}{Phys. Rev.} \textbf{\bibinfo{volume}{52}},
  \bibinfo{pages}{365} (\bibinfo{year}{1937}).

\bibitem[{\citenamefont{Abrikosov and
  Beneslavskii}(1970)}]{Abrikosov_Beneslavskii_1970}
\bibinfo{author}{\bibfnamefont{A.~A.} \bibnamefont{Abrikosov}}
  \bibnamefont{and} \bibinfo{author}{\bibfnamefont{S.~D.}
  \bibnamefont{Beneslavskii}}, \bibinfo{journal}{Zh. Eksp. Teor. Fiz.}
  \textbf{\bibinfo{volume}{59}}, \bibinfo{pages}{1280} (\bibinfo{year}{1970}),
  \bibinfo{note}{[Sov. Phys. JETP {\bf 32}, 699 (1971)]}.

\bibitem[{\citenamefont{Armitage et~al.}(2018)\citenamefont{Armitage, Mele, and
  Vishwanath}}]{Armitage_Mele_Vishwanath_2018}
\bibinfo{author}{\bibfnamefont{N.~P.} \bibnamefont{Armitage}},
  \bibinfo{author}{\bibfnamefont{E.~J.} \bibnamefont{Mele}}, \bibnamefont{and}
  \bibinfo{author}{\bibfnamefont{A.}~\bibnamefont{Vishwanath}},
  \bibinfo{journal}{Rev. Mod. Phys.} \textbf{\bibinfo{volume}{90}},
  \bibinfo{pages}{015001} (\bibinfo{year}{2018}).

\bibitem[{\citenamefont{Liu et~al.}(2017)\citenamefont{Liu, Hu, Graf, Cao,
  Radmanesh, Adams, Zhu, Cheng, Liu, Phelan et~al.}}]{Liu_et_al_2017}
\bibinfo{author}{\bibfnamefont{J.~Y.} \bibnamefont{Liu}},
  \bibinfo{author}{\bibfnamefont{J.}~\bibnamefont{Hu}},
  \bibinfo{author}{\bibfnamefont{D.}~\bibnamefont{Graf}},
  \bibinfo{author}{\bibfnamefont{H.~B.} \bibnamefont{Cao}},
  \bibinfo{author}{\bibfnamefont{S.~M.~A.} \bibnamefont{Radmanesh}},
  \bibinfo{author}{\bibfnamefont{D.~J.} \bibnamefont{Adams}},
  \bibinfo{author}{\bibfnamefont{Y.~L.} \bibnamefont{Zhu}},
  \bibinfo{author}{\bibfnamefont{G.~F.} \bibnamefont{Cheng}},
  \bibinfo{author}{\bibfnamefont{X.}~\bibnamefont{Liu}},
  \bibinfo{author}{\bibfnamefont{W.~A.} \bibnamefont{Phelan}},
  \bibnamefont{et~al.}, \bibinfo{journal}{Nature Materials}
  \textbf{\bibinfo{volume}{16}}, \bibinfo{pages}{905} (\bibinfo{year}{2017}).

\bibitem[{\citenamefont{Belitz et~al.}(1997)\citenamefont{Belitz, Kirkpatrick,
  and Vojta}}]{Belitz_Kirkpatrick_Vojta_1997}
\bibinfo{author}{\bibfnamefont{D.}~\bibnamefont{Belitz}},
  \bibinfo{author}{\bibfnamefont{T.~R.} \bibnamefont{Kirkpatrick}},
  \bibnamefont{and} \bibinfo{author}{\bibfnamefont{T.}~\bibnamefont{Vojta}},
  \bibinfo{journal}{Phys. Rev. B} \textbf{\bibinfo{volume}{55}},
  \bibinfo{pages}{9452} (\bibinfo{year}{1997}).

\bibitem[{\citenamefont{Belitz et~al.}(1999)\citenamefont{Belitz, Kirkpatrick,
  and Vojta}}]{Belitz_Kirkpatrick_Vojta_1999}
\bibinfo{author}{\bibfnamefont{D.}~\bibnamefont{Belitz}},
  \bibinfo{author}{\bibfnamefont{T.~R.} \bibnamefont{Kirkpatrick}},
  \bibnamefont{and} \bibinfo{author}{\bibfnamefont{T.}~\bibnamefont{Vojta}},
  \bibinfo{journal}{Phys. Rev. Lett.} \textbf{\bibinfo{volume}{82}},
  \bibinfo{pages}{4707} (\bibinfo{year}{1999}).

\bibitem[{\citenamefont{Kirkpatrick and
  Belitz}(2012)}]{Kirkpatrick_Belitz_2012b}
\bibinfo{author}{\bibfnamefont{T.~R.} \bibnamefont{Kirkpatrick}}
  \bibnamefont{and} \bibinfo{author}{\bibfnamefont{D.}~\bibnamefont{Belitz}},
  \bibinfo{journal}{Phys. Rev. B} \textbf{\bibinfo{volume}{85}},
  \bibinfo{pages}{134451} (\bibinfo{year}{2012}).

\bibitem[{\citenamefont{Brando et~al.}(2016)\citenamefont{Brando, Belitz,
  Grosche, and Kirkpatrick}}]{Brando_et_al_2016a}
\bibinfo{author}{\bibfnamefont{M.}~\bibnamefont{Brando}},
  \bibinfo{author}{\bibfnamefont{D.}~\bibnamefont{Belitz}},
  \bibinfo{author}{\bibfnamefont{F.~M.} \bibnamefont{Grosche}},
  \bibnamefont{and} \bibinfo{author}{\bibfnamefont{T.~R.}
  \bibnamefont{Kirkpatrick}}, \bibinfo{journal}{Rev. Mod. Phys.}
  \textbf{\bibinfo{volume}{88}}, \bibinfo{pages}{025006}
  (\bibinfo{year}{2016}).

\bibitem[{\citenamefont{Belitz et~al.}(2005)\citenamefont{Belitz, Kirkpatrick,
  and Rollb{\"u}hler}}]{Belitz_Kirkpatrick_Rollbuehler_2005}
\bibinfo{author}{\bibfnamefont{D.}~\bibnamefont{Belitz}},
  \bibinfo{author}{\bibfnamefont{T.~R.} \bibnamefont{Kirkpatrick}},
  \bibnamefont{and}
  \bibinfo{author}{\bibfnamefont{J.}~\bibnamefont{Rollb{\"u}hler}},
  \bibinfo{journal}{Phys. Rev. Lett.} \textbf{\bibinfo{volume}{94}},
  \bibinfo{pages}{247205} (\bibinfo{year}{2005}).

\bibitem[{\citenamefont{Abrikosov et~al.}(1963)\citenamefont{Abrikosov, Gorkov,
  and Dzyaloshinski}}]{Abrikosov_Gorkov_Dzyaloshinski_1963}
\bibinfo{author}{\bibfnamefont{A.~A.} \bibnamefont{Abrikosov}},
  \bibinfo{author}{\bibfnamefont{L.~P.} \bibnamefont{Gorkov}},
  \bibnamefont{and} \bibinfo{author}{\bibfnamefont{I.~E.}
  \bibnamefont{Dzyaloshinski}}, \emph{\bibinfo{title}{Methods of Quantum Field
  Theory in Statistical Physics}} (\bibinfo{publisher}{Dover, New York},
  \bibinfo{year}{1963}).

\bibitem[{\citenamefont{Wegner}(1979)}]{Wegner_1979}
\bibinfo{author}{\bibfnamefont{F.}~\bibnamefont{Wegner}}, \bibinfo{journal}{Z.
  Phys. B} \textbf{\bibinfo{volume}{35}}, \bibinfo{pages}{207}
  (\bibinfo{year}{1979}).

\bibitem[{\citenamefont{Belitz and
  Kirkpatrick}(2012)}]{Belitz_Kirkpatrick_2012a}
\bibinfo{author}{\bibfnamefont{D.}~\bibnamefont{Belitz}} \bibnamefont{and}
  \bibinfo{author}{\bibfnamefont{T.~R.} \bibnamefont{Kirkpatrick}},
  \bibinfo{journal}{Phys. Rev. B} \textbf{\bibinfo{volume}{85}},
  \bibinfo{pages}{125126} (\bibinfo{year}{2012}).

\bibitem[{\citenamefont{Kirkpatrick and
  Belitz}(2019)}]{Kirkpatrick_Belitz_2019}
\bibinfo{author}{\bibfnamefont{T.~R.} \bibnamefont{Kirkpatrick}}
  \bibnamefont{and} \bibinfo{author}{\bibfnamefont{D.}~\bibnamefont{Belitz}},
  \bibinfo{journal}{Phys. Rev. B} \textbf{\bibinfo{volume}{99}},
  \bibinfo{pages}{085109} (\bibinfo{year}{2019}).

\bibitem[{\citenamefont{Carneiro and Pethick}(1977)}]{Carneiro_Pethick_1977}
\bibinfo{author}{\bibfnamefont{G.~M.} \bibnamefont{Carneiro}} \bibnamefont{and}
  \bibinfo{author}{\bibfnamefont{C.~J.} \bibnamefont{Pethick}},
  \bibinfo{journal}{Phys. Rev. B} \textbf{\bibinfo{volume}{16}},
  \bibinfo{pages}{1933} (\bibinfo{year}{1977}).

\bibitem[{\citenamefont{Betouras et~al.}(2005)\citenamefont{Betouras, Efremov,
  and Chubukov}}]{Betouras_Efremov_Chubukov_2005}
\bibinfo{author}{\bibfnamefont{J.}~\bibnamefont{Betouras}},
  \bibinfo{author}{\bibfnamefont{D.}~\bibnamefont{Efremov}}, \bibnamefont{and}
  \bibinfo{author}{\bibfnamefont{A.}~\bibnamefont{Chubukov}},
  \bibinfo{journal}{Phys. Rev. B} \textbf{\bibinfo{volume}{72}},
  \bibinfo{pages}{115112} (\bibinfo{year}{2005}).

\bibitem[{pos()}]{positivity_footnote}
\bibinfo{note}{Fluctuations suppress the tendency of the fermion system to
  order ferromagnetically, and thus decrease $\chis^{(0)}$ compared to the
  free-fermion result. A magnetic field suppresses the fluctuations and thus
  leads to a positive correction to $\chis^{(0)}$. See
  Ref.~\onlinecite{Brando_et_al_2016a} for a discussion of this point.}

\bibitem[{int()}]{interaction_footnote}
\bibinfo{note}{The nonanalyticity is not present in a Fermi gas because an
  interaction is necessary to produce the mixing between positive and negative
  frequencies that is crucial for the coupling of the soft modes to the
  observables.}

\bibitem[{\citenamefont{Hertz}(1976)}]{Hertz_1976}
\bibinfo{author}{\bibfnamefont{J.}~\bibnamefont{Hertz}},
  \bibinfo{journal}{Phys. Rev. B} \textbf{\bibinfo{volume}{14}},
  \bibinfo{pages}{1165} (\bibinfo{year}{1976}).

\bibitem[{\citenamefont{Zhang et~al.}(2009)\citenamefont{Zhang, {C-X.~Liu},
  {X-L.~Qi}, Dai, Fang, and {S-C.~Zhang}}}]{Zhang_et_al_2009}
\bibinfo{author}{\bibfnamefont{H.}~\bibnamefont{Zhang}},
  \bibinfo{author}{\bibnamefont{{C-X.~Liu}}},
  \bibinfo{author}{\bibnamefont{{X-L.~Qi}}},
  \bibinfo{author}{\bibfnamefont{X.}~\bibnamefont{Dai}},
  \bibinfo{author}{\bibfnamefont{Z.}~\bibnamefont{Fang}}, \bibnamefont{and}
  \bibinfo{author}{\bibnamefont{{S-C.~Zhang}}}, \bibinfo{journal}{Nature Phys.}
  \textbf{\bibinfo{volume}{5}}, \bibinfo{pages}{438} (\bibinfo{year}{2009}).

\bibitem[{\citenamefont{Kotov et~al.}(2012)\citenamefont{Kotov, Uchoa, Pereira,
  Guinea, and {Castro Neto}}}]{Kotov_et_al_2012}
\bibinfo{author}{\bibfnamefont{V.~N.} \bibnamefont{Kotov}},
  \bibinfo{author}{\bibfnamefont{B.}~\bibnamefont{Uchoa}},
  \bibinfo{author}{\bibfnamefont{V.~M.} \bibnamefont{Pereira}},
  \bibinfo{author}{\bibfnamefont{F.}~\bibnamefont{Guinea}}, \bibnamefont{and}
  \bibinfo{author}{\bibfnamefont{A.~H.} \bibnamefont{{Castro Neto}}},
  \bibinfo{journal}{Rev. Mod. Phys.} \textbf{\bibinfo{volume}{84}},
  \bibinfo{pages}{1067} (\bibinfo{year}{2012}).

\bibitem[{\citenamefont{Wang et~al.}(2013)\citenamefont{Wang, Weng, Wu, Dai,
  and Fang}}]{Wang_et_al_2013}
\bibinfo{author}{\bibfnamefont{Z.}~\bibnamefont{Wang}},
  \bibinfo{author}{\bibfnamefont{H.}~\bibnamefont{Weng}},
  \bibinfo{author}{\bibfnamefont{Q.}~\bibnamefont{Wu}},
  \bibinfo{author}{\bibfnamefont{X.}~\bibnamefont{Dai}}, \bibnamefont{and}
  \bibinfo{author}{\bibfnamefont{Z.}~\bibnamefont{Fang}},
  \bibinfo{journal}{Phys. Rev. B} \textbf{\bibinfo{volume}{88}},
  \bibinfo{pages}{125427} (\bibinfo{year}{2013}).

\bibitem[{\citenamefont{Neupane et~al.}(2014)\citenamefont{Neupane, Xu, Sankar,
  Alidoust, Bian, Liu, Belopolski, Chang, Jeng, Lin
  et~al.}}]{Neupane_et_al_2014}
\bibinfo{author}{\bibfnamefont{M.}~\bibnamefont{Neupane}},
  \bibinfo{author}{\bibfnamefont{S.-Y.} \bibnamefont{Xu}},
  \bibinfo{author}{\bibfnamefont{R.}~\bibnamefont{Sankar}},
  \bibinfo{author}{\bibfnamefont{N.}~\bibnamefont{Alidoust}},
  \bibinfo{author}{\bibfnamefont{G.}~\bibnamefont{Bian}},
  \bibinfo{author}{\bibfnamefont{C.}~\bibnamefont{Liu}},
  \bibinfo{author}{\bibfnamefont{I.}~\bibnamefont{Belopolski}},
  \bibinfo{author}{\bibfnamefont{T.-R.} \bibnamefont{Chang}},
  \bibinfo{author}{\bibfnamefont{H.-T.} \bibnamefont{Jeng}},
  \bibinfo{author}{\bibfnamefont{H.}~\bibnamefont{Lin}}, \bibnamefont{et~al.},
  \bibinfo{journal}{Nature Commun.} \textbf{\bibinfo{volume}{5}},
  \bibinfo{pages}{3786} (\bibinfo{year}{2014}).

\bibitem[{cut()}]{cutoff_footnote}
\bibinfo{note}{As in the case of a vanishing spin-orbit interaction, $h$ gives
  the crucial soft modes a mass, and this is what leads to nonanalytic terms in
  the free-energy functional and its derivatives with respect to $h$. The mass
  provided by $h$ is thus very different from that provided by $v$, which
  simply renders certain modes irrelevant for the hydrodynamic properties of
  the system.}

\bibitem[{\citenamefont{Millis}(1993)}]{Millis_1993}
\bibinfo{author}{\bibfnamefont{A.~J.} \bibnamefont{Millis}},
  \bibinfo{journal}{Phys. Rev. B} \textbf{\bibinfo{volume}{48}},
  \bibinfo{pages}{7183} (\bibinfo{year}{1993}).

\bibitem[{\citenamefont{Conduit et~al.}(2009)\citenamefont{Conduit, Green, and
  Simons}}]{Conduit_Green_Simons_2009}
\bibinfo{author}{\bibfnamefont{G.~J.} \bibnamefont{Conduit}},
  \bibinfo{author}{\bibfnamefont{A.~G.} \bibnamefont{Green}}, \bibnamefont{and}
  \bibinfo{author}{\bibfnamefont{B.~D.} \bibnamefont{Simons}},
  \bibinfo{journal}{Phys. Rev. Lett.} \textbf{\bibinfo{volume}{103}},
  \bibinfo{pages}{207201} (\bibinfo{year}{2009}).

\bibitem[{\citenamefont{Karahasanovic et~al.}(2012)\citenamefont{Karahasanovic,
  Kr{\"u}ger, and Green}}]{Karahasanovic_Kruger_Green_2012}
\bibinfo{author}{\bibfnamefont{U.}~\bibnamefont{Karahasanovic}},
  \bibinfo{author}{\bibfnamefont{F.}~\bibnamefont{Kr{\"u}ger}},
  \bibnamefont{and} \bibinfo{author}{\bibfnamefont{A.~G.} \bibnamefont{Green}},
  \bibinfo{journal}{Phys. Rev. B} \textbf{\bibinfo{volume}{85}},
  \bibinfo{pages}{165111} (\bibinfo{year}{2012}).

\end{thebibliography}

\end{document}